\documentclass[runningheads]{llncs}
\usepackage[T1]{fontenc}
\usepackage{graphicx}
\usepackage{xcolor}
\AtBeginDocument{%
  \providecommand\BibTeX{{%
    \normalfont B\kern-0.5em{\scshape i\kern-0.25em b}\kern-0.8em\TeX}}}

\usepackage{bbm}
\usepackage{multirow}
\usepackage[inline]{enumitem}
\usepackage{graphicx}
\usepackage{sistyle}
\usepackage[ruled]{algorithm2e}
\usepackage{booktabs}
\usepackage{caption}
\SIthousandsep{,}
\usepackage{makecell}
\usepackage[skip=0pt]{caption}
\usepackage{amsmath}
\usepackage{amsfonts}
\usepackage{subfigure}
\usepackage{hyperref}
\usepackage{array} 
\usepackage{graphicx}
\usepackage{diagbox}
\usepackage{textcomp}
\usepackage[normalem]{ulem}
\usepackage[sort,numbers]{natbib}
\usepackage{acronym}

\acrodef{IR}{information retrieval}

\begin{document}

\title{On the Robustness of\\ Generative Information Retrieval Models:\\ An Out-of-Distribution Perspective}

\titlerunning{On the Robustness of Generative Information Retrieval Models}

\author{Yu-An Liu\inst{1,2}\orcidID{0000-0002-9125-5097} \and
Ruqing Zhang\inst{1,2}\orcidID{0000-0003-4294-2541} \and \\
Jiafeng Guo\inst{1,2}\thanks{Corresponding author}\orcidID{0000-0002-9509-8674} \and 
Changjiang Zhou\inst{1,2}\orcidID{0009-0000-0005-9465} \and
Maarten de Rijke\inst{3}\orcidID{0000-0002-1086-0202} \and
Xueqi Cheng\inst{1,2}\orcidID{0000-0002-5201-8195}
}

%
\authorrunning{Y. Liu et al.}
%
\institute{CAS Key Lab of Network Data Science and Technology, ICT, CAS \and
University of Chinese Academy of Sciences\and
University of Amsterdam\\
\email{\{liuyuan21b, zhangruqing, guojiafeng, zhouchangjiang23s, cxq\}@ict.ac.cn} \\
\email{m.derijke@uva.nl}
}

\maketitle

\begin{abstract}
Generative information retrieval methods retrieve documents by directly generating their identifiers. 
Much effort has been devoted to developing effective generative \ac{IR} models. 
Less attention has been paid to the robustness of these models. 
It is critical to assess the out-of-distribution (OOD) generalization of generative \ac{IR} models, i.e., how would such models generalize to new distributions? 
To answer this question, we focus on OOD scenarios from four perspectives in retrieval problems: (i)~\textit{query variations}; (ii)~\textit{unseen query types}; (iii)~\textit{unseen tasks}; and (iv)~\textit{corpus expansion}. 
Based on this taxonomy, we conduct empirical studies to analyze the OOD robustness of representative generative \ac{IR} models against dense retrieval models.
Our empirical results indicate that the OOD robustness of generative \ac{IR} models is in need of improvement
By inspecting the OOD robustness of generative \ac{IR} models we aim to contribute to the development of more reliable IR models.
The code is available at \url{https://github.com/Davion-Liu/GR_OOD}.
 
\keywords{Generative retrieval  \and Robustness \and Out-of-distribution.}
\end{abstract}

\acresetall

\section{Introduction}
With the development of representation learning techniques \cite{devlin2018bert}, considerable progress has been made in dense retrieval based on the ``index-retrieve'' pipeline \cite{guo2022semantic,chen2017efficient,matveeva2006high}. 
\Ac{IR} approaches based on the index-retrieve pipeline may suffer from a large memory footprint and difficulties in end-to-end optimization. 
Recently, a generative information retrieval paradigm has been proposed~\cite{metzler2021rethinking}. 
In this paradigm, different components for indexing and retrieval are fully parameterized with a single consolidated model. 
Specifically, a sequence-to-sequence (seq2seq) learning framework is employed to directly predict the identifiers of relevant documents (docids) with respect to a given query. 

Current research on generative \ac{IR} is often conducted in a homogeneous and narrow setting. 
That is, work on generative \ac{IR} often assumes that the training and test examples are independent and identically distributed (IID). 
Under the IID assumption, the generative \ac{IR} models that have been proposed so far have achieved promising performance on large-scale document retrieval tasks \cite{chen2022corpusbrain,tay2022transformer, wang2022neural}. 
However, in real-world scenarios, the IID assumption may not always be satisfied: the test distribution is usually unknown and possibly different from the training distribution. 
Put differently, high IID accuracy does not necessarily translate into out-of-distribution (OOD) robustness for document retrieval. 
Besides, pre-trained transformers, which usually serve as the backbone of existing generative \ac{IR} models, may rely on spurious cues and annotation artifacts are less likely to include OOD examples \cite{hendrycks2020pretrained}.
So far, little is known about the OOD robustness of generative \ac{IR} models. 

In this work, we systematically study OOD robustness across various families of retrieval models, including generative, dense, and sparse retrieval models.
In particular, we focus on comparing the robustness of generative \ac{IR} models with that of the other models.
We decompose OOD robustness into a model's generalization ability to 
\begin{enumerate*}[label=(\roman*)]
    \item query variations, 
    \item unseen query types, 
    \item unseen tasks, and 
    \item corpus expansion
\end{enumerate*}    
Each generalization ability perspective corresponds to a different OOD scenario.
Based on this taxonomy, we design corresponding experiments and conduct empirical studies to analyze the robustness of several representative generative \ac{IR} models against dense retrieval models. 

For our experiments, we employ the comprehensive knowledge-intensive language tasks (KILT) benchmark \cite{petroni-etal-2021-kilt}, which comprises eleven datasets across five KILT tasks. 
With its \emph{distinct tasks} and \emph{multiple corpora} for several of its tasks, KILT is ideal for an analysis of OOD robustness. 
In the future we will also try to explore more general datasets and models for experimentation.
In this work, following \cite{chen2022corpusbrain, deautoregressive}, we consider the retrieval task of KILT, in which the model should retrieve a set of Wikipedia pages as evidence for the final prediction with respect to the input query. 

Our experimental results reveal that, overall, generative \ac{IR} models perform poorly in terms of OOD robustness. Different generative \ac{IR} models display different types of generalizability performance in different OOD scenarios.
As a result, there is considerable scope for future robustness improvements. 
With our findings, we draw attention to an understudied research area. 

\section{Related Work}
\subsection{Sparse and dense retrieval models} 
Sparse retrieval models build representations of queries and documents based on the bag-of-words (BoW) assumption \cite{zhang2010understanding}, where each text is treated as a multiset of its words, ignoring grammar and word order \cite{guo2022semantic,StephenRobertson1994SomeSE}.
During the past decades, we have witnessed sparse retrieval models going through quick algorithmic shifts from early heuristic models \cite{salton1975vector},  vector space models \cite{salton1975vector}, to probabilistic models~\cite{StephenRobertson1994SomeSE,JayPonte1998ALM}.
BM25 \cite{robertson2009probabilistic}, as a representative of probabilistic models, is widely used for its efficiency while guaranteeing retrieval performance.

With the development of deep learning, many researchers
have turned to dense retrieval models \cite{Karpukhin2020DensePR,Khattab2020ColBERTEA,zhao2024dense}, which have been proven to be effective in capturing latent semantics and extracting effective features. 
Dense retrieval models typically adopt a bi-encoder architecture to encode queries and documents into low-dimension embeddings
and use embedding similarities as estimated relevance scores for effective retrieval~\cite{guo2022semantic}. 
\citet{Karpukhin2020DensePR} were pioneers in discovering that fine-tuning BERT to learn effective dense representations, called DPR, outperforms traditional retrieval methods like BM25.
Subsequently, researchers began exploring various fine-tuning techniques to enhance dense retrieval models, such as mining hard negatives~\cite{Xiong2021ApproximateNN,Zhan2021OptimizingDR}, late interaction~\cite{Khattab2020ColBERTEA}.
Recently, researchers have also investigated pre-training tasks for dense retrieval~\cite{gao2021condenser, ma2022contrastive}.
Although these methods greatly improve the performance of dense retrieval models, they follow the same bi-encoder architecture represented by the DPR and usually come with considerable memory demands and computational overheads.

\subsection{Generative \ac{IR} models}
Generative \ac{IR} has recently garnered increasing interest \cite{metzler2021rethinking,chen2022gere,wang2022neural,bevilacqua2022autoregressive}.
Generative \ac{IR} retrieves documents by directly generating their identifiers based on the given query.
It offers an end-to-end solution for document retrieval tasks \cite{metzler2021rethinking, chen-2023-unified} and allows for better exploitation of the capabilities of large generative language models.
For example, 
\citet{deautoregressive} proposed an autoregressive entity retrieval model and  
\citet{tay2022transformer} introduced a differentiable search index (DSI) and represent documents as atomic ids, naive string, or semantic strings.
\citet{chen2022corpusbrain} proposed a pre-trained generative \ac{IR} model called CorpusBrain to encode all information of the corpus within its parameters in a general way. 
Rather than using BART \cite{lewis2020bart} directly as a generative IR model, CorpusBrain can capture relevance signals within documents, leading to promising retrieval performance in a wide range of knowledge-intensive language tasks.
However, so far the robustness of generative \ac{IR} models has been overlooked by the community.

\subsection{Out-of-distribution in IR} 
OOD robustness refers to a model's ability to maintain performance when encountering data that differs from the distribution of the training data \cite{hendrycks2016baseline}. 
In real-world applications, retrieval models often face unseen data, highlighting the challenges of out-of-distribution robustness \cite{wu2022neural,thakur2beir, liu2025robust,liu2024robust,liu2024robust_survey}.
Current studies on OOD robustness in IR have their own limitations.
For example, \citet{wu2022neural} only explored the OOD generalization performance of neural ranking models.
Some work has been devoted to alleviating the poor performance of dense retrieval in the scenarios of query variants~\cite{zhuang2021dealing,penha2022evaluating,chen2022towards,sidiropoulos2022analysing} or zero/few-shot of corpus~\cite{yu2022coco,liang2020embedding,thakur2beir}.
In this work, we focus on the OOD generalizability of generative \ac{IR} models and compare them analytically with representative retrieval models from other families.

\section{IID Settings for the Retrieval Problem}

For a better understanding of the OOD setting for the retrieval problem, we first briefly introduce the IID setting for the retrieval problem. 

Formally, given a dataset $\mathcal{D} = \{(q_i, D, Y_i)\}^{n}_{i=1}$, where $q_i$ denotes a query, $D = \{d_{1}, d_{2}, \ldots, d_{N}\}$ represents the corpus, and $Y = \{r_1, r_2, \ldots, r_l\}$ indicates the corresponding relevance label of each document in $D$.
A total order exists among the relevance labels such that $r_l \succ r_{l-1} \succ \cdots \succ r_1$, where $\succ$ denotes the order relation. 
Each query $q_i$ is associated with a list of corresponding labels $\mathbf{y}_i = \{y_{i1}, y_{i2}, \ldots ,y_{i, N}\}$, where $N$ denotes the corpus size.

Traditionally, a retrieval model could be a term-based retrieval mode~\cite{StephenRobertson1994SomeSE,JayPonte1998ALM} or a dense retrieval model \cite{Karpukhin2020DensePR,ma2022contrastive}.
Recently, generative \ac{IR} models have emerged as another paradigm \cite{chen2022corpusbrain,tay2022transformer,deautoregressive}.
Although the paradigm is different, these retrieval models have the same formal definition regarding the retrieval task.
Without loss of generality, we use $f$ to denote the retrieval model.
We consider the retrieval model $f$ learned on the dataset $\mathcal{D}$, which is drawn from the training distribution $\mathcal{G}$. 
For retrieval we employ the learned model $f$ to generate a score for any query-document pair $(q,d)$, reflecting the relevance degree of  $d$ given $q$. 
This set-up allows us to produce a permutation $\pi(q_t, D, f)$ according to predicted scores.
Given an effectiveness evaluation metric $M$, retrieval models are typically evaluated by the average performance over the test queries under the IID setting, i.e.,
\begin{equation}
	\mathbb{E}_{(q_t,D,\mathbf{y}_t)\sim\mathcal{G}} M(\pi(q_t, D, f), \mathbf{y}_t),
\label{equ: mean_performance}
\end{equation} 
where $q_t,D$ and $\mathbf{y}_t$ denote the query, the corpus and the label in the test set, respectively. Specifically, the test samples are supposed to be drawn from the same distribution as $\mathcal{G}$. 

\section{OOD Settings for the Retrieval Problem}
In this work, we define the OOD robustness of retrieval models in four ways, i.e., in terms of query variations, unseen query types, unseen tasks, and corpus expansion. 
For query variations, the models are trained on the original dataset $\mathcal{D}$ and tested on the same dataset with query variations.
For unseen query types and unseen tasks, the models are trained on an original dataset $\mathcal{D}$ and tested on a new dataset $\mathcal{D}'$ with the same task as $\mathcal{D}$ and on $\tilde{\mathcal{D}}$ with a task that is different from $\mathcal{D}$, respectively.
For corpus expansion, the new dataset $\mathcal{D}^n$ that the models are tested on, is an expansion of the original dataset $\mathcal{D}$.

\subsection{Query variations}
The query variations refer to different expressions of the same information need.
Therefore, a query and its variations usually correspond to the same related document. 
This query-level OOD aims to analyze the model's generalizability across different query variations within the dataset.

Formally, suppose that the examples $q_t, D$ and $\mathbf{y}_t$ are drawn from the training distribution $\mathcal{G}$. 
We aim to evaluate the models' performance on the query OOD example. 
Specifically, the testing scenario of OOD generalizability on query variations is defined as
\begin{equation}
	\mathbb{E}_{(q_t, D,\mathbf{y}_t)\sim\mathcal{G}} M(\pi(G(q_t), D, f), \mathbf{y}_t),
\label{equ: defensive ability against query attack}
\end{equation} 
where $G(q_t)$ denotes the query variations generated by the generator $G$.

\subsection{Unseen query types}
The unseen query types scenario refers to unseen types of queries that are due to new types of information needs on the same task.
Due to the query-specific provenance, query distributions differ between one query set to another, every though they focus on the same task.
This query-type-level OOD aims to analyze the model's generalizability across different query types.

Formally, suppose that the new types of queries OOD examples $q_t'$ and relevance label $\mathbf{y}_t'$ are drawn from the new distribution $\mathcal{G}'_Q$ and come from dataset $\mathcal{D}'$. 
The corpus of datasets $\mathcal{D}$ and $\mathcal{D}'$ are consistent as $D$.
Specifically, the testing scenario of OOD generalizability on unseen query types is defined as 
\begin{equation}
	\mathbb{E}_{(q_t',D,\mathbf{y_t'})\sim\mathcal{G}'_Q} M(\pi(q_t', D, f), \mathbf{y}_t').
\label{equ: OOD generalizability on unseen type}
\end{equation}
Note that the training dataset $\mathcal{D}$ and the testing dataset $\mathcal{D}'$ with unseen query types come from the same task. 

\subsection{Unseen tasks}
The unseen tasks scenario refers to distribution shifts arising from task shifts.
In practice, a retrieval model is usually trained to focus on a specific task and model a particular relevance pattern.  
Therefore, it is essential to evaluate how well a retrieval model, trained on datasets of a given task, can generalize to datasets of new tasks.
This pair-level OOD aims to analyze a model's generalizability across different retrieval tasks.

Formally, suppose that the new task OOD examples $\tilde{q}_t$, $\tilde{D}$ and $\tilde{\mathbf{y}}_t$ are drawn from the new distribution $\tilde{\mathcal{G}_T}$ and come from dataset $\tilde{\mathcal{D}}$. 
Specifically, the testing scenario of OOD generalizability on unseen corpus is defined as 
\begin{equation}
\mathbb{E}_{(\tilde{q}_t,\tilde{D},\tilde{\mathbf{y}}_t)\sim\tilde{\mathcal{G}}_T} M(\pi(\tilde{q}_t, \tilde{D}, f), \tilde{\mathbf{y}}_t).
\label{equ: OOD generalizability on unseen task}
\end{equation}
Note that the training dataset $\mathcal{D}$ and the test dataset $\tilde{\mathcal{D}}$ belong to different tasks, respectively. 

\subsection{Corpus expansion}
The corpus expansion scenario refers to the scenario for the trained IR model to maintain its retrieval performance under continuously arriving new documents.
In reality, an IR corpus may expand as new documents continuously enter the system. 
Along with this, queries related to these new documents will also emerge. 
Therefore, it is important to evaluate the model's adaptability to these unseen documents. 
This pair-level OOD analysis is aimed at assessing a model's ability to generalize to an expanding corpus.

Formally, suppose that the corpus update examples $q_t^n$, $D^n$ and $\mathbf{y}_t^n$ come from corpus expansion $\mathcal{D}^n$. 
Specifically, the testing scenario of OOD generalizability on corpus expansion is defined as 
\begin{equation}
	\mathbb{E}_{(q_t,D,\mathbf{y}_t)\sim\mathcal{G}_T} M(\pi(q_t^n, D^n, f), \mathbf{y}_t^n),
\label{equ: OOD generalizability on unseen task}
\end{equation}
Note that the test dataset $\mathcal{D}^n$ is an expansion of the training dataset $\mathcal{D}$.

\section{Experimental Setup}
We introduce the experimental setup for analyzing OOD robustness.

\begin{table}[t]
    \centering
    \setlength\tabcolsep{6pt}
    \caption{Statistics of datasets in the KILT benchmark. `-' denotes that the task does not provide ground-truth documents in the training set.}
    \label{table:TASK STATEMENT}
    \begin{tabular}{lll@{}rr}
        \toprule
        \textbf{Task} & \textbf{Label} & \textbf{Dataset} & \textbf{Train. size} & \textbf{Dev. size}  \\
        \midrule
        Dialogue 
        & \textbf{WoW} & Wizard of Wikipedia~\cite{wow} & 63,734 & 3,054\\
        \midrule
        \multirow{3}{*}{Entity linking} 
        & \textbf{AY2} & AIDA CoNLL-YAGO~\cite{aida}  & 18,395 & 4,784 \\
        & \textbf{WnWi} & WNED-WIKI~\cite{wned}  & - & 3,396 \\
        & \textbf{WnCw} & WNED-CWEB~\cite{wned}  & - & 5,599 \\
        \midrule
        Fact checking
        & \textbf{FEV} & FEVER~\cite{fever} & 104,966 & 10,444 \\
        \midrule
        \multirow{4}{*}{Open domain QA} 
        & \textbf{NQ} & Natural Questions~\cite{nq}  & 87,372 & 2,837 \\
        & \textbf{HoPo} & HotpotQA~\cite{hotpotqa}  & 88,869 & 5,600 \\
        & \textbf{TQA} & TriviaQA~\cite{triviaqa}  & 61,844 & 5,359 \\
        & \textbf{ELI5} & ELI5~\cite{eli5}  & - & 1,507 \\
        \midrule
        \multirow{2}{*}{Slot filling} 
        & \textbf{T-REx} & T-REx~\cite{trex}  & 2,284,168 & 5,000 \\
        & \textbf{zsRE} & Zero Shot RE~\cite{zsre}  & 147,909 & 3,724 \\
        \bottomrule
    \end{tabular}
\end{table}

\subsection{Datasets}

For four OOD settings, we construct four benchmark datasets based on the KILT benchmark \cite{petroni-etal-2021-kilt} (see Table~\ref{table:TASK STATEMENT}). 
Due to the submission frequency limits of the online leaderboard, we used the performance on the dev set to evaluate model performance. 
In the following, we describe the details of the constructed datasets for evaluating the OOD generalizability on query variation, unseen query types, unseen tasks, and corpus expansion, respectively. 
\begin{itemize}[leftmargin=*]
    \item \textbf{Dataset for query variations}. We use the queries in Fever (FEV) and Natural Questions (NQ) to generate their variations, as all of the retrieval models perform relatively well on these datasets. Four generation strategies are considered \cite{penha2022evaluating} to perturb input queries, including (1) \textbf{Misspelling} for randomly substituting existing characters; (2) \textbf{Naturality} for removing all stop words; (3) \textbf{Order} for randomly exchanging positions of two words; and (4) \textbf{Paraphrasing} for replacing non-stop words according to the similarity of counter-fitted word embeddings \cite{NikolaMrki2016CounterfittingWV}.
Examples of the generated query variations are listed in Table \ref{table:query variation}.

\item \textbf{Dataset for unseen query types}. We use the datasets under the open-domain QA task which covers the largest number of datasets in the KILT.
There are three full datasets in open domain QA, i.e., NQ, HoPo, TQA. 
These datasets contain different topics and provenances, i.e., web search queries \cite{nq}, multi-hop questions \cite{hotpotqa}, and trivia questions \cite{triviaqa}.

\item \textbf{Dataset for unseen tasks}. We use 5 tasks from the KILT benchmark, namely, dialogue (Dial.), entity linking (EL), fact checking (FC), open domain question answering (QA),  and slot filling (SF).
For each task in the KILT, we mix every training and test set of all datasets under each task separately to create a new dataset for that task.

\item \textbf{Dataset for corpus expansion}.
To mimic corpus expansion, we randomly sample 60\% documents from the whole Wikipedia pages to serve as the initial corpus $D_0$ and leave the other 40\% Wikipedia pages as the incremental corpus $D_1$.
To construct the downstream KILT training set corresponding to $D_0$,
We filter the original KILT training set by retaining only those query-document pairs where all relevant articles in the corresponding provenance exclusively belong to $D_0$.
Similarly, to construct the test set $Q_0$ and $Q_1$ corresponding to $D_0$ and $D_1$, we first filter the original dev set by retaining only those query-document pairs where all relevant articles in the corresponding provenance exclusively belong to $D_0$, and then construct the filtered and remaining dataset as $Q_0$ and $Q_1$ respectively.

\end{itemize}

\begin{table}[t]
\centering
 \setlength\tabcolsep{15pt}
 \caption{Synthetic queries using variation generators.}
\begin{tabular}{ll} 
\toprule
Original query & who wrote most of the declaration of independence  \\
\midrule
Misspelling & who wr\textbf{eit} most of the declaration of independence  \\
Naturality & \textbf{\sout{who}} wrote  most \textbf{\sout{of}} \textbf{\sout{the}} declaration \textbf{\sout{of}} independence \\
Order & who \textbf{declaration} most of the \textbf{wrote} of independence \\
Paraphrasing & who \textbf{authored} most of the declaration of independence \\
\bottomrule
\end{tabular}
\label{table:query variation}
\end{table}

\subsection{Retrieval models}
We use representative samples of models from different families:
\begin{itemize}[nosep,leftmargin=*]
\item \textbf{BM25}~\cite{robertson2009probabilistic} is a representative sparse retrieval model that estimates the relevance based on term frequency, document length, and document frequency.
\item \textbf{DPR}~\cite{Karpukhin2020DensePR} is a representative dense retrieval model that uses dual-encoder architecture and is trained with in-batch negatives and a few hard negatives selected with BM25. 
\item \textbf{BART}~\cite{lewis2020bart} is a Seq2Seq model applicable for sequence generation tasks. 
Following \cite{deautoregressive,chen2022corpusbrain}, we extract the query-title pairs from each dataset and fine-tune the BART for generative retrieval.
\item \textbf{CorpusBrain}~\cite{chen2022corpusbrain} is a pre-trained generative \ac{IR} model for knowledge-intensive language tasks.
We fine-tune CorpusBrain on every specific downstream KILT task.
\end{itemize}

\subsection{Evaluation}
To measure the OOD generalizability of the retrieval models, following \cite{wu2022neural}, we use \textbf{DR$_\mathit{OOD}$} (\%) to evaluate the drop rate between the retrieval performance $P_\mathit{OOD}$ under the OOD setting 
and the retrieval performance $P_\mathit{IID}$ under the IID setting, defined as,
\begin{equation} 
\label{DROOD}
	DR_\mathit{OOD} = \frac{P_\mathit{OOD} - P_\mathit{IID}}{P_\mathit{IID}},
\end{equation} 
where $P_\mathit{IID}$ denotes the retrieval performance of the model trained on the  training set corresponding to the test set.
And $P_\mathit{OOD}$ denotes the retrieval performance of the model trained on the  training set that is out-of-distribution for the test set. 
The ranking model would be more robust with a higher DR$_\mathit{OOD}$. 

The effectiveness metric for evaluating the retrieval performance in KILT is usually defined as \textbf{R-precision} (\%), which is suggested in the official instructions and widely used in previous works on KILT \cite{deautoregressive, chen2022corpusbrain, chen2022gere}. 
R-precision is calculated as $\frac{r}{R}$, where $R$ is the number of Wikipedia pages inside each provenance set and $r$ is the number of relevant pages among the top-$R$ retrieved pages.

\section{Results}
We examine the empirical results in the IID setting and the three OOD settings sequentially: query variations, unseen query types, unseen tasks, and corpus expansion.

\begin{table}[t]
\centering
  \caption{R-precision (\%) for the page-level retrieval task on the KILT dev data.}
  \setlength\tabcolsep{4pt}
  \begin{tabular}{l  c c c c c c c c c }
  \toprule
  & 
  \multicolumn{1}{c}{\textbf{Dial.}} & 
  \multicolumn{1}{c}{\textbf{EL}} & 
  \multicolumn{1}{c}{\textbf{FC}} & 
  \multicolumn{3}{c}{\textbf{Open Domain QA}} & 
  \multicolumn{2}{c}{\textbf{Slot Filling}} & 
  \\ 
  \cmidrule(r){2-2}
  \cmidrule(r){3-3}
  \cmidrule(r){4-4}
  \cmidrule(r){5-7}
  \cmidrule(r){8-9}
  \textbf{Model}& 
  \textbf{WoW} & 
  \textbf{AY2} & 
  \textbf{FEV} & 
  \textbf{NQ} & \textbf{HoPo} & \textbf{TQA} & 
  \textbf{T-REx} & \textbf{zxRE} & 
  \textbf{Avg.} 
  \\ 
  \midrule
  BM25  & 27.5 & \phantom{1}3.5 & 50.1 & 25.8 & 44.0 & 29.4 & 58.6 & 66.4 & 38.2\\
  DPR  & 25.2 & \phantom{1}2.1 & 52.9 & 53.9 & 26.1 & 42.8 & 13.5 & 28.4 & 30.6\\
  BART & 50.7 & 90.1 & 79.6 & 48.9 & 41.6 & 64.4 & 74.4 & 94.3 & 68.0\\
  CorpusBrain & \textbf{55.0} & \textbf{90.7} & \textbf{81.4} & \textbf{57.6} & \textbf{50.7} & \textbf{70.9} & \textbf{75.7} & \textbf{97.6} & \textbf{72.5}\\
  \bottomrule
  \end{tabular}
  \label{table:Baseline}
\end{table}

\subsection{Overall IID results}
We compare the selected retrieval models on the KILT benchmark.
From Table \ref{table:Baseline}, we can observe that the generative \ac{IR} models significantly outperform sparse and dense retrieval models like BM25 and DPR across all the datasets, indicating that combining the retrieval components into a unified model benefits effective corpus indexing.
CorpusBrain consistently outperforms BART on all five tasks, demonstrating that the adequately well-designed pre-training tasks for generative retrieval contribute to improving document understanding for generative \ac{IR} models.

\begin{table*}[t]
\centering
   \caption{R-precision / DR$_{OOD}$ for query variations on the FEV and NQ dev data. Significant performance degradation with respect to the corresponding  IID setting is denoted as `$-$' ($\textit{p-value} \leq 0.05$).}
   \setlength\tabcolsep{4pt}
  	\begin{tabular}{l  c c c c c}
  \toprule
  \textbf{Model} & \textbf{Original} & \textbf{Misspelling} & \textbf{Naturality} & \textbf{Order} & \textbf{Paraphrasing} \\
\midrule
  & \multicolumn{5}{c}{\textbf{FEV}} \\
       \midrule
BM25 &  50.1 & \textbf{31.8}/-36.5 & \textbf{49.2}/-0.02 & 22.3/\phantom{-0}0\phantom{.0} & 39.5/-21.2 \\
DPR  &  52.9 & 24.1/-54.4 & 32.4/-38.8 & 22.3/-57.8 & 34.8/-34.2 \\
BART  & 79.6 & 20.7/-74.0 & 38.3/-51.9 & 22.1/-72.2 & 34.7/-56.4 \\
CorpusBrain  & \textbf{81.4} & 26.0/-68.0 & 41.8/--48.6 & \textbf{27.7}/-66.0 & \textbf{40.6}/-50.1 \\
\midrule
  & \multicolumn{5}{c}{\textbf{NQ}} \\
\midrule
BM25  & 25.8 & 20.5/-20.5 & 25.4/-0.02 & 31.0/\phantom{-0}0\phantom{.0} & 22.1/-14.3 \\
DPR  & 53.9 & 25.6/-52.5 & 31.8/-41.0 & 31.0/-42.5 & 44.6/-17.3 \\
BART & 48.9 & 26.2/-46.4 & 39.1/-20.0 & 32.8/-32.9 & 43.4/-11.2 \\
CorpusBrain & \textbf{57.6} & \textbf{28.1}/-51.2 & \textbf{39.2}/-31.9 &  \textbf{36.1}/-37.3 & \textbf{50.1}/-13.0 \\

\bottomrule
    \end{tabular}
   \label{table: query ood}
\end{table*}

\subsection{Analysis of OOD generalizability on query variations}
Firstly, from Table \ref{table: query ood}, we provide a comprehensive performance analysis of all the retrieval models.
We can observe that some types of query variants, such as naturality and order, have little impact on BM25.
This is because BM25 uses bags of words to model documents and is insensitive to changes in word order and stop words.
Beyond that, there is a significant effectiveness drop for query variations in all dense and generative retrieval models. 
The results indicate that both dense retrieval models, as well as generative \ac{IR} models, are not robust to query variations, which complements the findings from previous work \cite{penha2022evaluating}.

When we compare the generalizability of the dense and generative \ac{IR} models, we find that the generative \ac{IR} models perform particularly poorly on \textbf{Misspelling} and \textbf{Order}.
One possible explanation would be that the generative \ac{IR} models generate document identifiers autoregressively based on the query, so query quality and word order greatly impact the generation effect.
When we look at the performance of generative \ac{IR} models, we can find that the CorpusBrain has better R-precision than BART, indicating that pre-training tasks tailored for generative retrieval help the model adapt better to query variations.

\begin{table}[t]
\centering
   \caption{R-precision/DR$_\mathit{OOD}$ for unseen query types on the open domain QA dataset of KILT. }
   \setlength\tabcolsep{13pt}
  	\begin{tabular}{l l c c c c c }
  \toprule
  & & \multicolumn{3}{c}{\bf Testing}
  \\
  \cmidrule{3-5}
  \textbf{Model} & \bf Training & \textbf{NQ} & \textbf{HoPo} & \textbf{TQA} \\ 
  \midrule
    \multirow{3}{*}{BM25} 
  & NQ & 25.8 & - & - \\ 
  & HoPo & - & 44.0 & - \\
  & TQA & - & - & 29.4 \\ \hline
  \multirow{3}{*}{DPR} 
  & NQ & 53.9 & 23.1/-11.5 & 29.2/-31.8 \\ 
  & HoPo & 41.2/-23.6 & 26.1 & 26.3/-38.6 \\
  & TQA & 42.3/-21.5 & 21.8/-16.5 & 42.8 \\ \hline
  \multirow{3}{*}{BART} 
  & NQ & 48.9 & 36.4/-12.5 & 50.7/-21.3 \\ 
  & HoPo & 18.8/-61.6 & 41.6 & 46.8/-27.3 \\
  & TQA & 26.5/-45.8 & 35.2/-15.4 & 64.4 \\ \hline
  \multirow{3}{*}{CorpusBrain} 
  & NQ & \textbf{57.6} & 47.0/\phantom{1}-7.3 & 52.7/-25.7\\ 
  & HoPo & 33.4/-42.0 & \textbf{50.7} & 48.6/-31.5 \\
  & TQA & 32.9/-42.9 & 44.7/-11.8 & \textbf{70.9} \\
  \bottomrule
  \end{tabular}
  \label{table: cross dataset}
\end{table}

\subsection{Analysis of OOD generalizability on unseen query types}
The results of OOD generalizability on unseen query types are shown in Table \ref{table: cross dataset}.
Note that BM25 does not rely on the training set, so its test results remain consistent across datasets.
When we look at the overall performance of all the dense and generative retrieval models, we can observe that as the shift of query types distributions, the performance of all models decreases significantly.
For DPR, some of the query types in which it had an advantage are instead inferior to BM25 in OOD scenarios.
This suggests that, even under the same task, neural retrieval models face challenges of poor OOD generalizability.
Consequently, it is important to consider the OOD performance for these unseen query types.

Comparing the performance of dense and generative \ac{IR} models, we find that generative \ac{IR} models exhibit worse generalizability on web search queries in the NQ dataset.
It indicates that, in terms of generalizability performance on unseen query types, generative \ac{IR} models behave differently and merit separate studies.
Furthermore, we observe that CorpusBrain demonstrates better generalizability than BART on unseen query types.
This could be attributed to the pre-training process of CorpusBrain, which effectively encodes relevant information for a given corpus to cope with potentially unknown queries, thereby enhancing its stability when encountering unseen query types.

\begin{table}[t]
\centering
   \caption{R-precision/DR$_\mathit{OOD}$ for unseen tasks on the 5 KILT task-mixed datasets.}
   \setlength\tabcolsep{3.5pt}
  	\begin{tabular}{l l c c c c c c }
  \toprule
  & & \multicolumn{5}{c}{\bf Testing}
  \\
  \cmidrule{3-7}
  \textbf{Model} & \textbf{Training} & \textbf{Dial.} & \textbf{EL}  & \textbf{FC} & \textbf{QA} & \textbf{SF} \\ 
\midrule
\multirow{5}{*}{BM25} 
  & Dial. & 27.5 & - & - & - & - \\ 
  & EL & - & \phantom{1}3.5 & - & - & - \\
  & FC & - & - & 50.1 & - & - \\ 
  & QA & - & - & - & 34.6 & - \\
  & SF & - & - & - & - & 61.9 \\
\midrule
\multirow{5}{*}{DPR} 
  & Dial. & 25.5 & \phantom{1}0.7/-66.7 & 48.2/\phantom{1}-8.9 & 25.1/-30.1 & 10.3/-46.6 \\ 
  & EL & 13.6/-46.7 & \phantom{1}2.1 & 46.8/-11.5 & 15.9/-55.7 & 13.6/-29.5 \\
  & FC & 24.0/\phantom{1}-5.9 & \phantom{1}0.8/-61.9 & 52.9 & 22.3/-37.9 & 15.2/-21.1 \\ 
  & QA & 22.5/-11.8 & \phantom{1}0.6/-71.4 & 50.8/\phantom{1}-4.0 & 35.9 & 12.1/-37.3 \\
  & SF & 10.0/-60.8 & \phantom{1}0.2/-90.5 & 45.4/-14.2 & 20.2/-43.7 & 19.3 \\
\midrule
\multirow{5}{*}{BART} 
  & Dial. & 49.7 & \phantom{1}8.6/-90.5 & 71.8/\phantom{1}-9.8 & 40.2/-31.8 & 69.2/-17.5 \\ 
  & EL & 21.5/-56.7 & 90.1 & 68.6/-13.8 & 24.6/-58.3 & 77.4/\phantom{1}-7.7 \\
  & FC & 48.1/\phantom{1}-3.2 & 10.0/-88.9 & 79.6 & 41.5/-29.7 & 81.1/\phantom{1}-3.3 \\ 
  & QA & 45.0/\phantom{1}-9.5 & \phantom{1}9.8/-89.1 & 76.4/\phantom{1}-4.0 & 59.0 & 77.6/\phantom{1}-7.5 \\
  & SF & 17.1/-65.6 & \phantom{1}3.7/-95.9 & 65.6/-17.6 & 36.3/-38.5 & 83.9 \\
\midrule
\multirow{5}{*}{CorpusBrain} 
   & Dial. & \textbf{58.0} & \phantom{1}6.9/-92.4 & 74.1/\phantom{1}-9.0 & 49.3/-18.8 & 77.7/\phantom{1}-7.8 \\
  & EL & 33.0/-43.1 & \textbf{90.7} & 68.6/-15.7 & 38.4/-36.7 & 62.7/-25.6 \\
  & FC & 46.8/-19.3 & \phantom{1}9.3/-89.7 & \textbf{81.4} & 48.9/-19.4 & 82.2/\phantom{1}-2.5 \\ 
  & QA & 46.3/-20.1 & \phantom{1}8.1/-91.1 & 79.2/\phantom{1}-2.7 & \textbf{60.7} & 78.7/\phantom{1}-6.6 \\
  & SF & 25.3/-56.4 & \phantom{1}4.9/-94.6 & 68.1/-16.3 & 43.7/-28.0 & \textbf{84.3} \\
  \bottomrule
  \end{tabular}
  \label{table: cross task}
\end{table}

\subsection{Analysis of OOD generalizability on unseen tasks}
Examining the overall performance of all retrieval models in Table \ref{table: cross task}, we observe that the generalizability defects for unseen tasks are common among models.
In the entity linking (EL) task, the models' generalization performance drops significantly, likely due to the task's distinct format compared to the others.

DPR lags behind BM25 almost across the board when faced with unseen tasks.
The reason may be the large differences in data distribution across tasks.
The semantic representations that DPR learns by learning from the original task are empirical and difficult to flexibly migrate to the new task.
While dense retrieval models have excellent performance, there are situations where traditional sparse retrieval models are rather more to be relied upon.

When we observe the performance between dense and generative \ac{IR} models, we find that, in general, generative \ac{IR} models have higher $DR_{OOD}$ on slot-filling (SF) task.
This could be because the format of this downstream task aligns with the pre-training tasks of the backbone generative models.
Comparing BART and CorpusBrain from the generative \ac{IR} models, we observe that CorpusBrain outperforms BART in most (13 out of 20) unseen task scenarios.
This may be attributed to the pre-training tasks of CorpusBrain.
CorpusBrain includes three tasks: Inner Sentence Selection (ISS), Lead Paragraph Selection (LPS), and Hyperlink Identifier Prediction (HIP). 
ISS models the semantic granularity differences between queries and documents in various retrieval requirements, helping to bridge the gap between different downstream tasks. 
This finding is consistent with the original analysis of CorpusBrain \cite{chen2022corpusbrain}.

\begin{table}[t]
\centering
   \caption{R-precision/DR$_\mathit{OOD}$ for corpus expansion on the 5 KILT task-specific datasets.}
 \renewcommand{\arraystretch}{0.85}
   \setlength\tabcolsep{4pt}
  	\begin{tabular}{l c c c c c c c }
  \toprule
   \textbf{Model} &  \textbf{Session} & \textbf{Dial.} & \textbf{EL} & \textbf{FC} & \textbf{QA} & \textbf{SF} \\ 
       \midrule
\multirow{2}{*}{BM25} 
  & \textbf{D$_0$} & 26.0 & \phantom{1}2.6 & 46.5 & 40.2 & 55.7 \\ 
  & \textbf{D$_1$} & 20.5/-21.2 & \phantom{1}1.8/-30.8 & 37.8/\phantom{1}-6.0 & 28.2/-29.9 & 46.7/-16.2 \\ 
\midrule
\multirow{2}{*}{DPR} 
  & \textbf{D$_0$} & 49.2 & \phantom{1}2.3 & 73.2  & 46.5 & 40.1 \\ 
  & \textbf{D$_1$}  & 28.7/-41.7 & \phantom{1}1.6/-30.4 & 65.7/-10.2 & 41.8/-10.1 & 35.0/-12.7 \\ 
\midrule
\multirow{2}{*}{BART} 
  & \textbf{D$_0$} & 42.7 &63.0 &74.5 &22.7 &63.6 \\ 
  & \textbf{D$_1$} & 47.0/\phantom{-}10.1 & 57.2/\phantom{1}-9.2 &70.7/\phantom{1}-5.1 &20.9/\phantom{1}-7.9 &47.0/-26.1 \\ 
\midrule
\multirow{2}{*}{CorpusBrain}
    & \textbf{D$_0$} & 36.4 & 64.0 & 78.8 & 41.2 & 81.2 \\ 
  & \textbf{D$_1$} & 15.1/-58.5 & 43.5/-32.0 & 56.7/-28.0 & 28.3/-31.3 & 74.7/\phantom{1}-8.0 \\ 
\bottomrule
  \end{tabular}
  \label{table: corpus expansion}
\end{table}

\subsection{Analysis of OOD generalizability on corpus expansion}
The result of OOD generalizability on corpus expansion is shown in Table \ref{table: corpus expansion}.
From the result, we can observe that, BM25 has an average ability to maintain retrieval performance.
When faced with incremental documents entering the corpus, the performance degradation of DPR is not significant. 
The possible reason for this is that under the same task, the newly arrived document belongs to the same topic as the old one, and DPR can build the complete semantic representation space from the original training.

For the generative IR model, both BART and CorpusBrain perform significantly better in corpus expansion than dense retrieval models like DPR.
Even with a lower DR$_\mathit{OOD}$ than BM25 and DPR, the overall performance of the generative IR model is still higher than that of them.
The reason for this is that generative retrieval uses a prefix tree to store indexes, and unseen indexes will still be distributed in the neighborhood of the indexes they are related to.
That is, generative IR models can extensively probe for relevant documents in the corpus through beam search, which underpins their generalization capabilities.

\section{Conclusion}

In this paper, we have analyzed the out-of-distribution robustness of several representative generative and dense retrieval models on the KILT benchmark. 
Specifically, we have proposed four perspectives to define out-of-distribution robustness. 
Our results exposed significant vulnerabilities in OOD robustness of generative \ac{IR} models.

We believe that the understanding of different forms of retrieval models can open up ideas from a robustness perspective.
As we observed, the dense retrieval model and the generative retrieval model perform differently for different OOD scenarios.
While there is some prior work that relates generative and dense retrieval~\citep{nguyen-2023-generative,wu-2024-generative}, the robustness perspective on their connection is missing -- what can we learn from their relative strengths and weaknesses to develop more robust retrieval models?

Concerning limitations presented here, we chose CorpusBrain and BART, which perform well on KILT, as representatives of generative IR models.
In future work, we will introduce more generative \ac{IR} models with more datasets to further explore the OOD robustness.
Due to inheriting the vulnerabilities of neural network models, neural IR models are also susceptible to being deceived by out-of-distribution adversarial examples \cite{liu2023black,liu2023topic,liu2024multi,liu2024perturbation,liu2025attachain}.
Future work should consider introducing new web search datasets into the benchmark to simulate more broader and potentially even more challenging OOD environments like adversarial attacks.
Our work highlights the need to create benchmarks that include various OOD perspectives to better understand the generative \ac{IR} models' robustness.

Finally, we will consider different docid forms of generative IR models to explore the differences in robustness performance between generative IR models.

\section*{Acknowledgements}
This work was funded by the National Natural Science Foundation of China (NSFC) under Grants No. 62472408,  the Strategic Priority Research Program of the CAS under Grants No. XDB0680102, the National Key Research and Development Program of China under Grants No. 2023YFA1011602, the Lenovo-CAS Joint Lab Youth Scientist Project, and the project under Grants No. JCKY2022130C039.
This work was also (partially) funded by the Dutch Research Council (NWO), under project numbers 024.004.022, NWA.1389.20.\-183, and KICH3.LTP.20.006, and the European Union's Horizon Europe program under grant agreement No. 101070212.

All content represents the opinion of the authors, which is not necessarily shared or endorsed by their respective employers and/or sponsors.

\clearpage

\bibliographystyle{splncs04nat}
\bibliography{references}

\end{document}